\documentclass[twocolumn,aps,prb,superscriptaddress]{revtex4-1}
\usepackage[latin1]{inputenc}
\usepackage{amsmath}
\usepackage{amsfonts}
\usepackage{amssymb}
\usepackage{graphicx}
\usepackage{epstopdf}
\usepackage{color}
\usepackage[version=4]{mhchem}
\newcommand{\degree}{\ensuremath{^\circ}}

\begin{document}
\title{Tuning the antiferromagnetic helical pitch length and nanoscale domain size in \ce{Fe3PO4O3} by magnetic dilution}
\author{Tarne, M.J.}
\affiliation{Department of Chemistry, Colorado State University, Fort Collins, Colorado 80523, USA}
\author{Bordelon, M.M.}
\affiliation{Department of Chemistry, Colorado State University, Fort Collins, Colorado 80523, USA}
\author{Calder, S.}
\affiliation{Quantum Condensed Matter Division, Oak Ridge National Laboratory, Oak Ridge, Tennessee 37831, USA}
\author{Neilson, J.R.}
\affiliation{Department of Chemistry, Colorado State University, Fort Collins, Colorado 80523, USA}
\author{Ross, K.A.}
\email{kate.ross@colostate.edu}
\thanks{Corresponding author}
\affiliation{Department of Physics, Colorado State University, Fort Collins, Colorado 80523, USA}

\begin{abstract}
The insulating magnetic material \ce{Fe3PO4O3} features a non-centrosymmetric lattice composed of \ce{Fe^{3+}} triangular units.  Frustration, due to competing near neighbor ($J_1$) and next nearest neighbor ($J_2$) antiferromagnetic interactions, was recently suggested to be the origin of an antiferromagnetic helical ground state with unusual needle-like nanoscale magnetic domains in \ce{Fe3PO4O3}.   Magnetic dilution is shown here to tune the ratio of these magnetic interactions, thus providing deeper insight into this unconventional antiferromagnet.  Dilution of the \ce{Fe^{3+}} lattice in \ce{Fe3PO4O3} was accomplished by substituting non-magnetic \ce{Ga^{3+}} to form the solid solution series \ce{Fe_{3-x}Ga_xPO4O3} with $x = 0.012, 0.06, 0.25, 0.5, 1.0, 1.5$.   Magnetic susceptibility and neutron powder diffraction data from this series are presented.  A continuous decrease of the both the helical pitch length and the domain size is observed with increasing dilution up to at least $x = 0.25$, while for $x \ge 0.5$, the compounds lack long range magnetic order entirely.  The decrease in the helical pitch length with increasing $x$ can be qualitatively understood by reduction of the ratio of $J_2/J_1$ in the Heisenberg model, consistent with mean field considerations. Intriguingly, the magnetic correlation length in the $ab$ plane remains nearly equal to the pitch length for each value of $x \le 0.25$, showing that the two quantities are intrinsically connected in this unusual antiferromagnet. 
\end{abstract}
\maketitle

\section{Introduction}

Magnetic frustration,  arising from competing interactions, can lead to a variety of interesting phenomena.  The roster of frustration-induced effects includes noncollinear spin structures\cite{Yoshimori1959,Kaplan1959}, spin glasses\cite{Itoh1994}, multiferroics\cite{Blake2005}, and spin liquids\cite{balents2010spin}. In each case, the delicate balance of interactions means that small perturbations, such as magnetic dilution, can lead to dramatic effects.  

When frustration leads to helical magnetic structures, the possibility of the formation of Skyrmions, or other topological objects such as vortices or magnetic bubbles, arises \cite{okubo2012multiple}.   Skyrmions in particular are of current interest for their possible uses in spintronics applications\cite{Sampaio2013}.  They can form a thermodynamically stable Skyrmion-lattice phase\cite{Adams2012, Mnzer2010, Pfleiderer2009, Nagaosa2013} (also describable as a multi-$k$ magnetic structure), or they can be generated as single defects by combining helical magnetic domain walls \cite{Zhou2014,Zhang2016}. The presently-known materials forming Skyrmions are based on the competition of Dzyaloshinskii-Moriya (DM) interactions and ferromagnetic (FM) Heisenberg exchange, forming locally \emph{ferromagnetic} or \emph{ferrimagnetic} helical states, from which the Skyrmion lattice arises in finite magnetic field.  The local FM background makes them susceptible to the Skyrmion Hall effect, which hinders their efficient manipulation in devices\cite{Barker2016}.  

Frustrated antiferromagnets have been proposed as an alternative material class which are capable of generating of Skyrmions\cite{okubo2012multiple, leonov2015multiply, hayami2016bubble}. The presence of antiferromagnetic (AFM) interactions may act to stabilize AFM Skyrmions, which are predicted to avoid the Skyrmion Hall effect \cite{Barker2016, Zhang2016}.  AFM Skyrmions have yet to be observed experimentally.   In order to generate AFM Skyrmions in a real material, one starting point may be the identification of an AFM helical structure with a high density of domain walls.  This scenario would suggest an underlying instability towards the formation of topological defects in the helical AFM state, and also provides the possibility of creating interesting spin textures from domain wall intersections.  In this light, the frustrated antiferromagnet \ce{Fe3PO4O3} is of prime interest \cite{Ross2015}.    

\begin{figure*}
	\centering
	\includegraphics[width=0.9\linewidth]{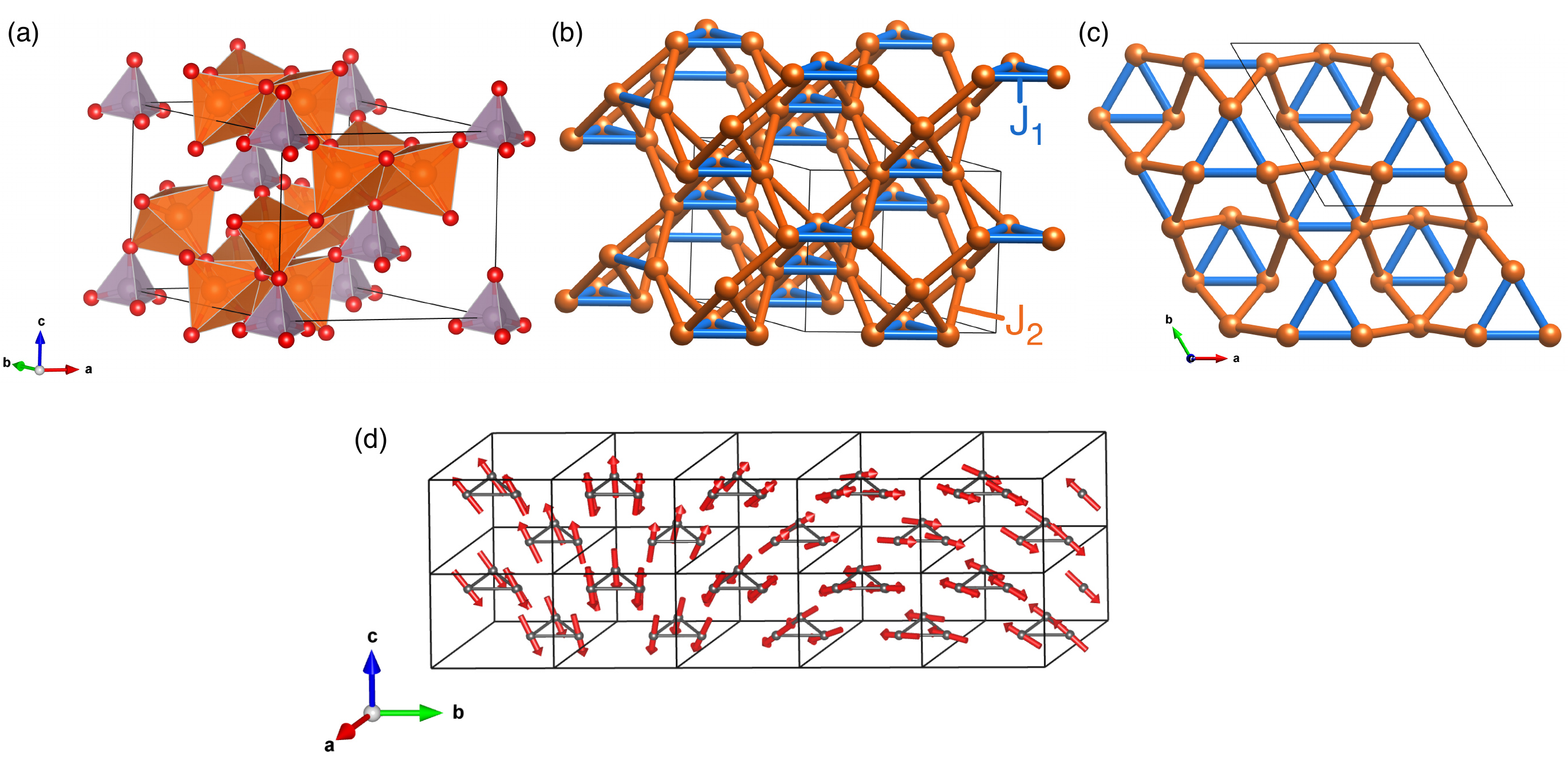}
	\caption{a) Crystal structure of \ce{Fe3PO4O3}. Iron is shown in orange, phosphorus in purple, and oxygen in red. b) Fe-Fe magnetic interactions in \ce{Fe3PO4O3 }. $J_1$ bonds are shown in blue, $J_2$ bonds are shown in orange. c) View along the $c$ axis, highlighting the triangular Fe$^{3+}$ units in \ce{Fe3PO4O3}.  d) Antiferromagnetic helical structure for the parent compound, Fe$_3$PO$_4$O$_3$.  As the magnetic sublattice is diluted (increasing $x$ in Fe$_{3-x}$Ga$_x$PO$_4$O$_3$), the same type of structure persists, but the helical pitch length and $ab$ plane correlation length decrease in tandem.}
	\label{fig:J2_J1_figure}
\end{figure*}

The structure of \ce{Fe3PO4O3} (noncentrosymmetric space group $R3m$ with room temperature lattice parameters of $a=b= 8.006$ \AA \ and $c = 6.863$ \AA \ in the hexagonal setting) can be described as triangular units of distorted \ce{FeO5} trigonal bipyramids coordinated by phosphate groups as shown in Figure \ref{fig:J2_J1_figure}a. Figures  \ref{fig:J2_J1_figure}b and c show the connectivity of magnetic \ce{Fe^{3+} } cations; $J_1$ nearest-neighbor interactions are shown in blue within the triangles and $J_2$ next-nearest-neighbor interactions are shown in orange between neighboring triangles. When these interactions are AFM and isotropic (as expected in \ce{Fe3PO4O3} based on the lack of single-ion anisotropy expected for $S = \frac{5}{2}$ \ce{Fe^{3+}} and significant direct exchange interactions\cite{Gavoille1987}), this model produces helical magnetic states near the frustrated point $J_2/J_1 \approx 2$.  Magnetization\cite{Modaressi1983, Gavoille1987}, specific heat\cite{Shi2013}, M\"{o}ssbauer spectroscopy\cite{Modaressi1983, Gavoille1987}, and neutron powder diffraction\cite{Gavoille1987,Ross2015} all confirm that \ce{Fe3PO4O3} orders into an AFM structure below $T_N$ = 163 K.  The magnetic structure of \ce{Fe3PO4O3} was recently shown to be quite unusual, however, in that the AFM helical state (Figure \ref{fig:J2_J1_figure} d) is only partially ordered; the correlation length of the magnetic order extends to at least 90 nm (900 \AA)  along the $c$ axis, but remains limited to $\xi_{ab} <$ 10 nm in the \textit{ab} plane down to at least $T=4$ K \cite{Ross2015}, suggesting a high density of domain walls separating helically-ordered domains.   As shown in our previous work, frustration is likely to be responsible for the orientation of the domain walls (which run parallel to $c$, producing needle-like domains)\cite{Ross2015}, but the origin of the high density of these defects is presently unknown.  Here, we shed some light on this issue by showing that the defect spacing is intimately tied to the helical pitch length in \ce{Fe3PO4O3}, which we show can be varied by greater than a factor of two through dilution of the magnetic lattice. 

%We previously reported magnetization measurements on \ce{Fe_3PO4O3} and the magnetically diluted series \ce{Fe_{3-x}Ga_xPO4O3} and obtained accurate estimates of the mean-field Curie-Weiss interaction strength, $\theta_{CW}$,  which ranges from $\theta_{CW} < -1000$ K for \ce{Fe3PO4O3} to --300 K for \ce{Fe_{0.5}Ga_{2.5}PO4O3} \cite{Ross2015}. These large negative values imply strong AFM interactions, even in the presence of significant magnetic dilution.

%\begin{figure*}
%	\includegraphics[width=0.65\linewidth]{magnetic_structure2.png}
%	\caption{Antiferromagnetic helical structure for the parent compound, Fe$_3$PO$_4$O$_3$, as was determined in Ref. \onlinecite{Ross2015}.  As the magnetic sublattice is diluted (increasing $x$ in Fe$_{3-x}$Ga$_x$PO$_4$O$_3$), the same type of structure persists, but the helical pitch length and $ab$ plane correlation length decrease in tandem. }
%	\label{fig:mag_struct}
%\end{figure*}

\begin{figure}[b]
	\centering
	\includegraphics[width=0.95\linewidth]{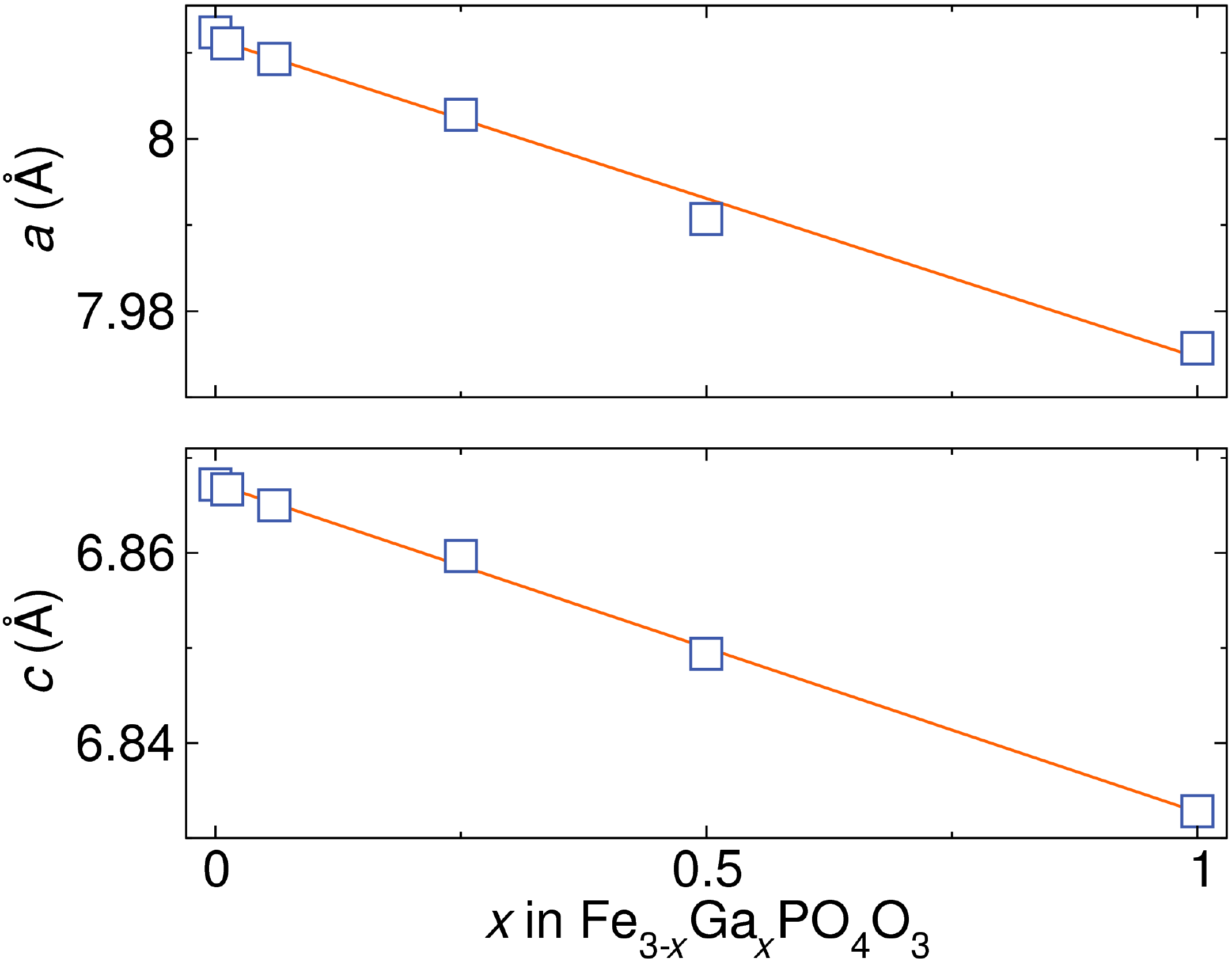}
	\caption{Extracted lattice parameters, $a$ (top panel) and $c$ (bottom panel) from Rietveld refinements against  room-temperature laboratory X-ray diffraction data, collected using NaCl as an internal standard.  The orange lines are the interpolation between the previously reported values for \ce{Fe3PO4O3}\cite{Ross2015} and \ce{Ga3PO4O3}\cite{boudin1998ga3po7}.
}
	\label{fig:lattice_params}
\end{figure}

\begin{figure*}
	\includegraphics[width=0.65\linewidth]{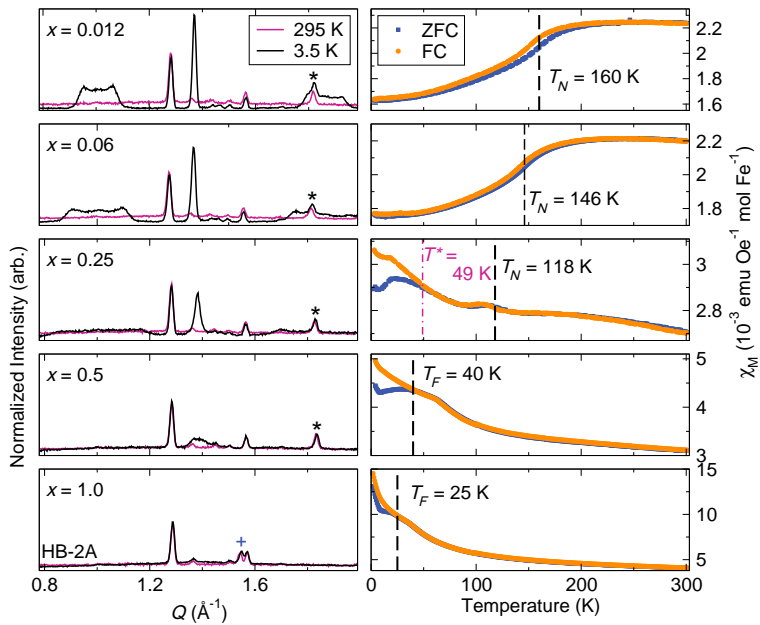}
	\caption{
		Neutron powder diffraction (left) and magnetic susceptibility (right) for $x$ = 0.012, 0.06, 0.25, 0.5, and 1.0 in \ce{Fe_{3-x}Ga_xPO4O3}. The magnetic Bragg reflections centered at $Q$ = 1, 1.35, 1.75 \AA$^{-1}$ decrease in intensity and broaden as gallium content increases. The magnetic susceptibility shows a decrease in the magnetic transition temperature with increased gallium substitution and eventually the development of a ZFC/FC splitting suggestive of a spin-frozen state. Although LRO order exists in $x$ = 0.25 at 118 K, there is a lower temperature transition at 49 K denoted as $T^{*}$ that does not correspond to changes in the static properties of the magnetic order. Small quantities of \ce{FePO4} impurities in the NPD data are shown as black stars and \ce{GaPO4} impurities as blue addition signs.}
	\label{fig:susceptibility_all}
\end{figure*}

In this contribution we examine the magnetic order in gallium-doped \ce{Fe_{3-x}Ga_xPO_4O_3} through low temperature magnetic susceptibility and neutron powder diffraction. We use the helical magnetic structure that was previously proposed \cite{Ross2015} (Fig. \ref{fig:J2_J1_figure} d) to model the neutron powder diffraction patterns of the gallium-substituted materials. For low dilution, $0 \leq x \leq 0.25$, the helical pitch length decreases with increasing $x$, which can be reproduced using a reduced ratio of $J_2/ J_1$ in the Heisenberg exchange model.  This is consistent with a mean field picture, since twice as many $J_2$ interactions are removed as compared to $J_1$ interactions for every Fe atom that is replaced with non-magnetic Ga. The size of the magnetic domains in the $ab$ plane, determined via the broadening of magnetic Bragg peaks, is also reduced with increasing $x$.  Curiously, the correlation length in the $ab$ plane (equal to half of the typical domain size) appears to be constrained to be nearly equal to the pitch length.   At higher values of gallium substitution, $x \geq 0.5$, long-range magnetic order in any direction is not observed, providing an estimate for the percolation threshold for this lattice.

\section{Methods}

Polycrystalline samples in the solid solution series \ce{Fe_{3-x}Ga_xPO4O3} were prepared as previously described\cite{Ross2015}. \ce{Fe_2O_3} and \ce{Ga2O3} were dried overnight at 600\degree C and \ce{FePO_4\cdot $n$H2O} was converted to anhydrous \ce{FePO4} by heating in air at 900\degree C overnight in alumina crucibles, and then stored in a desiccator. Powder X-ray diffraction data were collected using a NaCl internal standard to check for purity and for shifts in the lattice parameter.  Data were collected using a Bruker D8 Discover DaVinci Powder X-ray Diffractometer.  Rietveld refinements including explicit modeling of the NaCl standard were performed using the TOPAS software package.   The refinements confirm a solid-solution behavior of \ce{Fe_{3-x}Ga_xPO4O3}.    The extracted lattice parameters follow Vegard's law for a solid solution between \ce{Fe3PO4O3} and \ce{Ga3PO4O3}; the observed lattice parameters agree well with the  extrapolated lattice parameters for a given value of $x$, suggesting that the actual composition reflects the nominal composition (Figure~\ref{fig:lattice_params}).   

Magnetization measurements were performed using a vibrating sample magnetometer in a Quantum Design Physical Properties Measurement System.  The samples were measured in an applied field of $\mu_0\textrm{H = 1 T}$ from $T=$ 1.8 K to 300 K after cooling in the absence of an applied magnetic field (zero field cooled measurement, a.k.a. ZFC), and then measured from 300 to 1.8 K in a 1T field (field cooled measurement, a.k.a. FC).

\begin{figure*}
	\centering
	\includegraphics[width=0.65\linewidth]{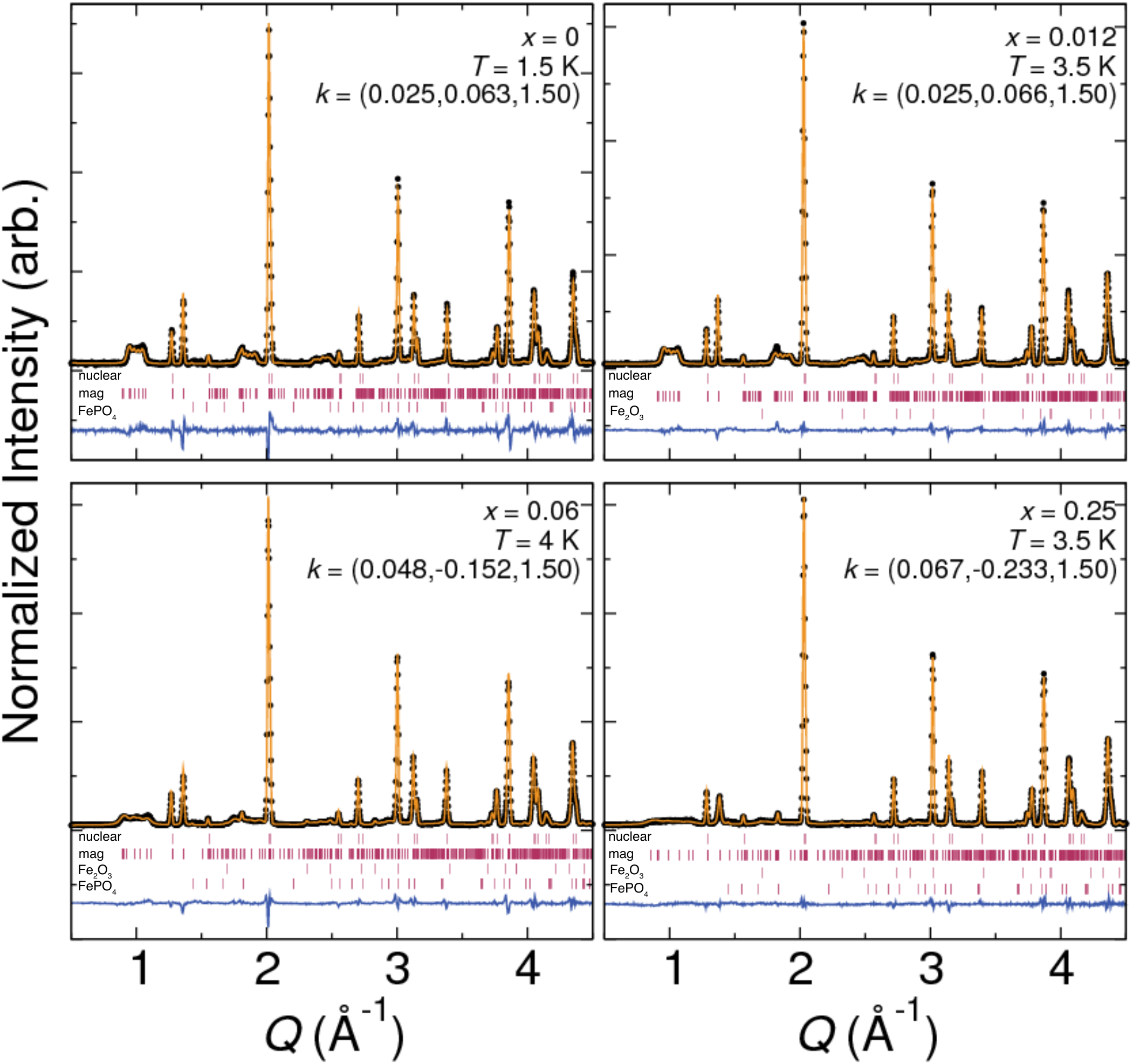}
	\caption{Rietveld refinements using the Fullprof software package for the magnetically ordered members of the series \ce{Fe_{3-x}Ga_xPO4O3}. The data are shown as black points, the fit is in orange, and tick marks show the central peak positions of nuclear, magnetic, and impurity phases. The difference curve is in blue.}
	\label{fig:all_refinements}
\end{figure*}

Neutron powder diffraction (NPD) measurements were carried out on 2.3-2.6 g polycrystalline samples of \ce{Fe_{3-x}Ga_xPO4O3} with $x$ = 0, 0.012, 0.06, 0.25, 0.5, 1.0, and 1.5 using beamline HB-2A at the High Flux Isotope Reactor at Oak Ridge National Laboratory. The samples were measured at temperatures ranging from 3.5 to 295 K using a helium closed cycle refrigerator; collection temperatures were selected based on features observed in susceptibility measurements. The instrument provided a constant wavelength incident neutron beam with $\lambda$ = 2.4123 \AA \ from the Ge(113) monochromator, and 21 arcminute collimators were placed between the monochromator and the sample, while 12 arcminute collimators were used between the sample and each of the 44 detectors.  This configuration produces a resolution of $\Delta Q /Q$ = $1.5\times 10^{-2}$ at $Q$ = 1 \AA$^{-1}$.\cite{garlea2010high} Magnetic and nuclear Rietveld refinements were performed using the FullProf software package\cite{RodrguezCarvajal1993}.

Magnetic structures were determined for the classical $J_1-J_2$ Heisenberg Hamiltonian for various ratios of $J_2/J_1$ using a numerical energy minimization routine implemented in the program SpinW (a simplex method with periodic boundary conditions)\cite{Toth2015}.

\section{Results}
The low-temperature magnetic susceptibility as well as neutron diffraction patterns at high and low temperatures is shown in Figure \ref{fig:susceptibility_all} for all $x>0$. A slight splitting between the FC and ZFC curves in the parent compound ($x = 0$, shown in Ref. \onlinecite{Ross2015}) begins slightly above $T_N$ and persists down to at least 2 K.  Similar features are observed for $x=0.012$ and $x=0.06$ (Fig. \ref{fig:susceptibility_all}), although $T_N$ decreases with increasing $x$. $T_N$ is estimated at 160 K for $x=0.012$ and 146 K for $x=0.06$, based on the location of the the largest positive $dM/dT$ in the ZFC susceptibility, a feature which is coincident with a peak in the specific heat for $x=0$ as shown in Ref. \onlinecite{Ross2015}.  

A more dramatic ZFC/FC splitting occurs for $x > 0.06$.  Based on comparison to the temperature dependent neutron powder diffraction patterns (Appendix, Fig. \ref{fig:neutron_raw}), discussed next, this splitting occurs at the temperature $T^*$ which is \emph{below} $T_N$ for $x=0.25$, possibly indicating a freezing of ``free spins'' produced by dilution (i.e., spins which are liberated from strong constraints on their orientation due to missing neighbors, also known as ``orphan spins'' \cite{schiffer1997two}). For $x>0.5$, a long-range ordered state is not observed, and the ZFC/FC splitting in the susceptibility might signal spin freezing throughout the entire sample, as in a spin glass. Accordingly, the temperature of the ZFC/FC splitting is denoted as $T_F$ for $x>0.5$.  The overall magnitude of the susceptibility increases for $x>0.5$, consistent with moving toward the paramagnetic limit as dilution acts to suppress the strong AFM correlations known to be present even up to 900 K in Fe$_3$PO$_4$O$_3$ \cite{Ross2015}.

The magnetic structures for $x>0$ remain qualitatively similar to the parent compound, as confirmed by NPD (Figure \ref{fig:all_refinements}), although $T_N$ decreases with gallium substitution as expected for overall weaker interactions produced by dilution \cite{Martinez1992,Fak2005}. The eventual disappearance of long-range order, which we characterize by absence of a well-formed magnetic peak near $Q$=1.35 \AA$^{-1}$, occurs for $x \ge 0.5$. 

\begin{figure}
	\includegraphics[width=0.9\linewidth]{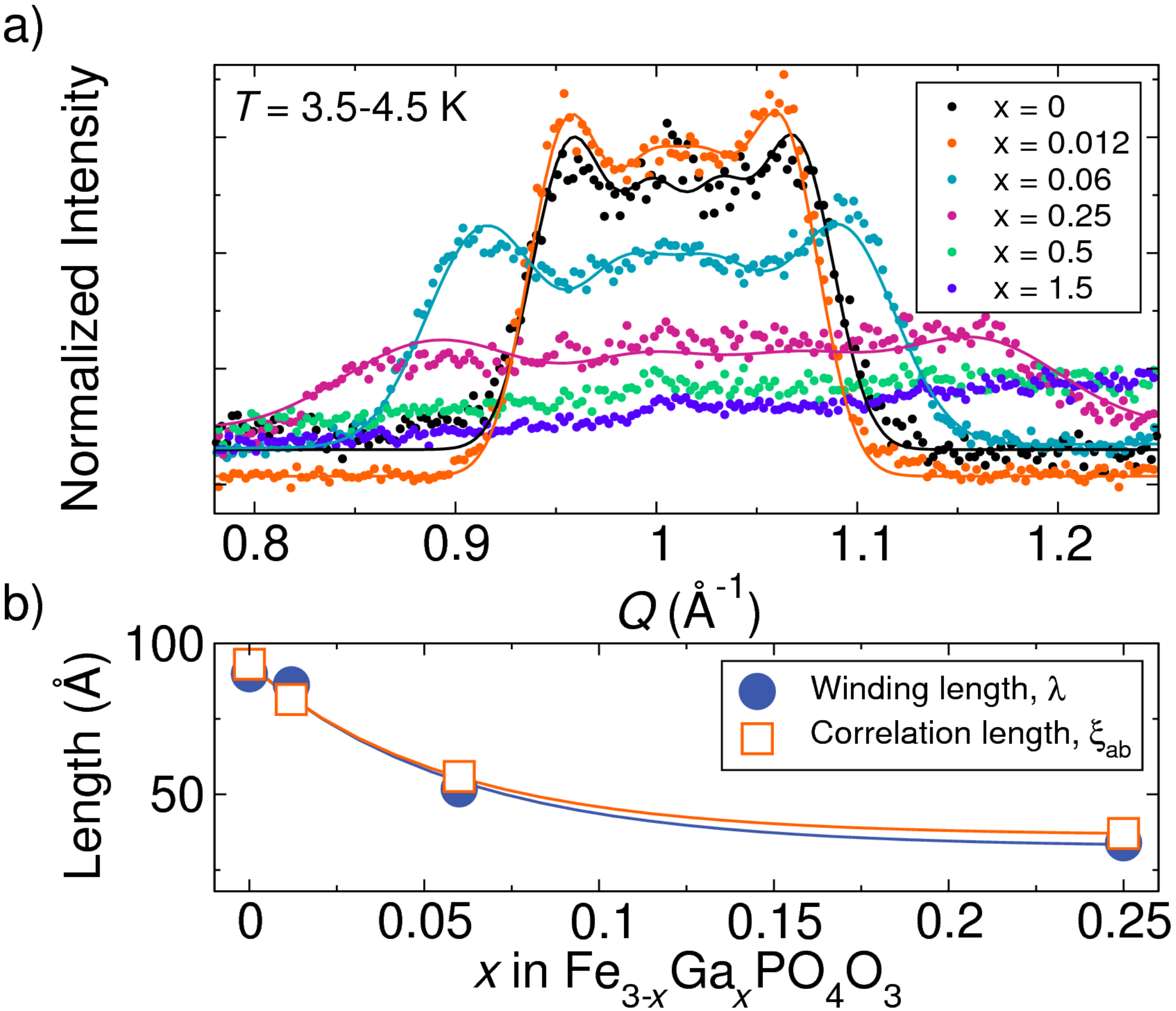}
	\caption{a) Broad magnetic peak measured by NPD for different gallium concentrations in  \ce{Fe_{3-x}Ga_xPO4O3} at $T =$ 3.5 - 4.5 K; data is shown as points and the profile-matched fits are solid lines (see main text). The broad, flat-topped magnetic Bragg peak broadens and decreases in intensity with increased gallium content, consistent with a decrease in $J_2$ mean-field interaction strength relative to $J_1$.  b) The helical winding length and $ab$ plane magnetic correlation length are shown in the bottom panel as a function of gallium content; both decrease with gallium substitution and track each other. Error bars are smaller than the symbols. Solid lines added to guide the eye.}
	\label{fig:horsetooth_all}
\end{figure}

The helical magnetic structure model previously used for the parent compound, with an ordering wavevector $\textbf{k}_h = (\delta_a,\delta_b,1.5)$ (hexagonal setting of $R3m$) and needle-like finite size broadening indicating a magnetic domain size in the $ab$ plane, was fit to the data for $x \leq 0.25$ (Figure \ref{fig:all_refinements}). The magnitude of this incommensuration, $\lvert\delta\rvert = \sqrt{(\delta_a + \frac{1}{2}\delta_b)^2 + (\frac{\sqrt{3}}{2}\delta_b)^2}$, increases with $x$, implying a reduced helical pitch length $l = \frac{a}{|\delta|}$ as shown in Figure \ref{fig:horsetooth_all}b. The incommensuration is constant in temperature below $T_N$ for each sample. For $x=0.5$ at $T = 3.5$ K, the broad magnetic peaks are not visible and the normally sharp $(\delta_a,\delta_b,3/2)$ peak near $Q \sim 1.35$ \AA$^{-1}$ is significantly broadened. For samples with $x \geq 1.0$, there are no magnetic Bragg features except for diffuse scattering centered at $Q \sim 1.35$ \AA$^{-1}$ as shown in Figures \ref{fig:susceptibility_all} and \ref{fig:horsetooth_all}.

%The $T_N$ decreases with gallium substitution, as deduced from both susceptibility and NPD data. The Bragg reflections from magnetic order disappears between 75 K and 140 K for $x = 0.25$, and for $x = 0.5$ the broad, flat-topped Bragg reflections are not observed, consistent with a spin-freezing transition in the dc susceptibility, Figure \ref{fig:susceptibility_all}. This decrease in $T_N$ is often observed in magnetically diluted materials\cite{Martinez1992,Fak2005} and can be attributed to the decrease in mean interaction strength.

%The full width at half maximum (FWHM) of the 6 broadened satellite peaks comprising the magnetic feature near 1 \AA$^{-1}$ was taken as the convolution of the instrument resolution and the Gaussian broadening. 

Correlation lengths can be extracted from the magnitude of the peak broadening required to match the slopes of the edges of the broad magnetic features.  The coexistence of a sharp, resolution limited magnetic Bragg peak (near $Q \sim 1.35$ \AA$^{-1}$) and broad, flat topped peaks (such as that near $Q \sim 1$ \AA$^{-1}$) can be described by anisotropic broadening, which affects peaks that have large $\vec{Q}$ components in the $ab$ plane (i.e., produced by needle-shaped domains).  The Lorentzian broadening employed in the Fullprof Rietveld refinement software package does not accurately capture the shape of the broad peaks, particularly the slopes of the peak edges.  In order to extract the $ab$-plane correlation length more accurately, we used a custom-made analysis script which fits the broad peak near 1 \AA$^{-1}$ and the sharp peak near 1.35 \AA$^{-1}$ simultaneously, applying a Gaussian broadening beyond the instrument resolution to the former, but not the latter.  For this analysis, the peak locations were set by a single magnetic propagation vector, but intensities were permitted to refine independently (i.e., a LeBail, also known as profile-matching, pattern refinement method was used).  This method allows us to refine the ordering wavevector, which is strongly constrained by the central $Q$ of the sharp peak, and the correlation length $\xi_{ab}$ simultaneously.  The fits are shown in Figure \ref{fig:horsetooth_all}, while the fits to the whole NPD pattern obtained using Lorentzian broadening as implemented in Fullprof are shown in Figure \ref{fig:all_refinements}.  The correlation length is extracted as $\xi_{ab} = \pi/\beta$, where $\beta$ is the full width at half maximum (FWHM) of the satellite peaks with $Q$ being the independent variable (the correlation length is taken to be the radius of the Scherrer domain size, assuming a shape factor of K = 1 \cite{hammond2001basics}).   The $ab$ plane correlation length decreases with increasing $x$, and remains approximately equal to the helical pitch length for all values of $x$ (Figure \ref{fig:horsetooth_all}). The marked increase in domain wall density (from $\frac{1}{\pi\xi_{ab}^2} = 0.0037$ nm$^{-2}$ at $x = 0$ to 0.023 nm$^{-2}$ at $x = 0.25$) shows that magnetic dilution greatly stabilizes the formation of domain walls.  Furthermore, the domain wall density is not consistent with a simple picture of nucleation at the gallium defects; assuming an even distribution of the defects, and taking $\frac{x}{3} \times 9$ Ga atoms per 55.5 \AA$^{2}$ (there are 9 sites in the unit cell, which has an $ab$ plane area of 55.5 \AA$^{2}$), gives a domain wall density of 1.3 nm$^{-2}$ at $x=$0.25, which is much larger than observed.

\section{Discussion}
The magnetic frustration in \ce{Fe3PO4O3} stems from the competition between the antiferromagnetic intra-triangle $J_1$ and inter-triangle $J_2$ interactions. As non-magnetic gallium substitutes into the structure, the magnetic $J_2$ and $J_1$ interactions are disrupted. For each gallium atom, two $J_1$ bonds and four $J_2$ bonds are broken, from which we expect a reduction in the average ratio $J_2/J_1$. The effect of this rebalancing of the exchange interactions was examined using a numerical minimization of the classical spin Hamiltonian,

\begin{equation}
\mathcal{H} = \sum_{<i,j>} J_1 \mathbf{S}_i\cdot\mathbf{S}_j + \sum_{<<i,j>>} J_2 \mathbf{S}_i\cdot\mathbf{S}_j
\label{eq:H}
\end{equation}

where the first sum is over nearest neighbors and the second is over next nearest neighbors.  

\begin{figure}
	\includegraphics[width=0.9\linewidth]{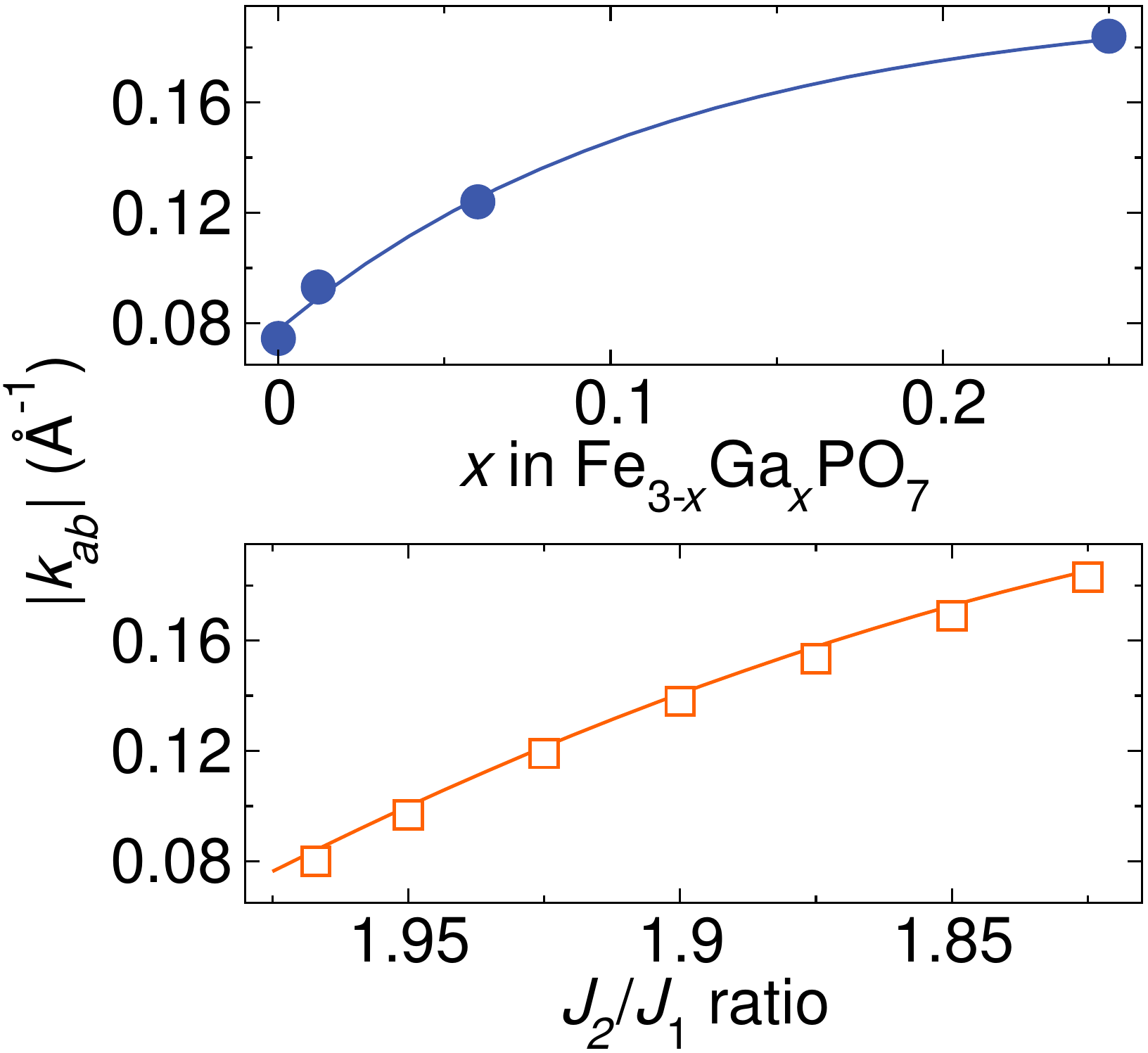}
	\caption{Experimental (top) and calculated (bottom) values of the length of the magnetic propagation vector in the $ab$ plane, $|k_{ab}|$, for Ga-doped \ce{Fe3PO4O3}. The mean-field model of decreased $J_2/J_1$ ratio with increased gallium substitution is consistent with experimental results. Lines added to guide the eye.}
	\label{fig:kab_exp_calc}
\end{figure}

Planar helical magnetic structures with the $\mathbf{k}_h = (\delta_a,\delta_b,1.5)$ ordering wavevector are produced when $J_2/J_1  \approx 2$.    The ratio of $J_2/J_1$ was varied slightly, and the resulting helical modulation ($|k_{ab}| = \frac{2\pi}{a}|\delta|$) was determined.  The variation in the helical pitch of the magnetic structure agrees qualitatively with the experimental results: $|k_{ab}|$ increases with decreasing $J_2/J_1$, as shown in Figure \ref{fig:kab_exp_calc}.  Matching the calculated $|k_{ab}|$ to that of the parent compound ($|k_{ab}| = 0.073$ \AA$^{-1}$), this approximation gives $J_2/J_1  = 1.96$ for $x = 0$ which is reduced to $J_2/J_1 = 1.84$ for $x = 0.25$, indicating that the mean $J_2$ interaction strength is reduced relative to $J_1$, as expected.

Some simplifying assumptions were made in the choice of the magnetic structure, as well as the Heisenberg model, discussed above.  To within the constraints of our powder averaged neutron diffraction data, we cannot distinguish between coplanar (helical) versus some non-coplanar (conical) magnetic structures, as discussed in Ref. \onlinecite{Ross2015}; we thus chose the former for simplicity.  Furthermore, the crystal structure lacks any inversion center, implying the presence of Dzyaloshinskii-Moriya (DM) interactions for every pair of Fe atoms, which may favor non-coplanar conical structures.   However, the simple model presented here captures the main features of the NPD of \ce{Fe3PO4O3} and its variation with magnetic dilution;  the pitch length and correlation length would not depend on the choice of a helical vs. conical structure, and although the mapping of $|k_{ab}|$ to the ratio of $J_2/J_1$ might be modified by the inclusion of DM interactions, we expect this to be a small effect since the DM interactions arise from spin orbit coupling, which is weak for Fe$^{3+}$.

%The neutron powder diffraction data of the solid solution series also lends insight into the unusually broad magnetic reflections. In our previous contribution, the peaks were modeled as a single-wavevector structure with Gaussian broadening due to finite domains in the \textit{ab} plane\cite{Ross2015}. In addition, models with multiple wavevectors, a square-wave incommensurate structure, and a conical structure were tested, each with less satisfactory modelling of the magnetic peak shape. We note that the broad peaks are more symmetric than expected from the Fe$^{3+}$ magnetic form factor and a single ordinary wavevector, and this becomes even more pronounced with the significantly broader peaks such as those in \ce{Fe_{2.75}Ga_{0.25}PO4O3}. This can be seen in Figure \ref{fig:horsetooth_all} in the difference between the single-wavevector model (solid line) and experimental data (points); the broad peaks in the model still have significant internal structure even with Gaussian broadening from the finite domains and the several symmetry-related wavevectors. Any attempt to add further broadening, either Lorentzian or Gaussian, results in less satisfactory modeling of the slopes on the peak sides. 

 We note that the intensity distribution of the broad peaks are more symmetric than expected, since the Fe$^{3+}$ magnetic form factor should tend to suppress the intensity at the high $Q$ side.  This disagreement becomes even more pronounced with the significantly broader peaks such as those in \ce{Fe_{2.75}Ga_{0.25}PO4O3}.  The inferred covalency of the Fe-O bonds in \ce{Fe3PO4O3}, discussed in Ref. \onlinecite{Ross2015}, would produce a less localized spin distribution, weakening the magnetic form factor.  Note that the profile-matching calculation shown in Figure \ref{fig:horsetooth_all} does not include the form factor, but the Fullprof refinements shown in Figure \ref{fig:all_refinements} do.   
% the intensities of the magnetic satellites were freely refined (as in a LeBail profile-matching type fit \REF) in order to extract the broadening. 

An unusual feature of the parent compound which extends to the the weakly diluted variants is the presence of needle-like magnetic domains, manifested as the broadening of magnetic peaks which have large $\vec{Q}$ components in the \textit{ab} plane.  Figure \ref{fig:horsetooth_all} shows that the correlation length, and hence domain size, decreases with gallium content.  Since magnetostriction has been proposed as an origin for the formation of antiferromagnetic domains\cite{minakov1990magnetostriction, Gomonay2002}, one might infer that the presence of smaller Ga ions at Fe sites creates strain or chemical pressure that locally increases the influence of magnetostriction, i.e., domain wall nucleation at impurities.  However, as noted above, assuming a uniform distribution of Ga does not reproduce the inferred domain wall density.  Some insight on the origin of the small domain size in the $ab$ plane is provided by Figure \ref{fig:horsetooth_all}b, which reveals that the correlation length and the helical winding length track each other for all measured values of $x$ below the percolation limit ($x \sim 0.5$).  Since the decrease in the winding length can be understood as a decrease in $J_2/J_1$, an intrinsic effect, the connection between the domain size and the pitch length intriguingly suggests that helical domains are truncated due to an intrinsic instability of the helical structure toward defect formation.   

%Speculating on this point further, one might imagine that the truncation of the helical structure occurs due to the formation of line-like defects from the superposition of different symmetry-related $k$-vectors, either due the intersection of domain walls, or alternatively, due to the magnetic structure being a multi-$k$ version of the helical state proposed here \footnote{As usual, one cannot distinguish between multi-k magnetic structures and domains of a single-k structure in \ce{Fe3PO4O3} from our NPD data}.  This occurs, for instance, in materials that form Skyrmion lattices (multi-$k$ helical structures), where the spacing between Skyrmions (line ``defects'')  is equal to the helical pitch length \cite{pfleiderer2010skyrmion,seki2012observation}.    However, if antiferromagnetic Skyrmions exist in \ce{Fe3PO4O3}, it cannot be that they form a long-range ordered Skyrmion lattice, as this would lead to resolution-limited satellite peaks, as in MnSi \cite{muhlbauer2009skyrmion}.  The short correlation length in the $ab$ plane observed in \ce{Fe3PO4O3} might instead imply something closer to a Skyrmion glass \cite{reichhardt2015collective}.  Further insight on the possible connection between the unusual short range correlated helical magnetic state in \ce{Fe3PO4O3} and a Skyrmion state would require the combination of real space measurements and neutron scattering measurements on single crystal samples.

\section{Conclusions}

Our study of the magnetic dilution series \ce{Fe_{3-x}Ga_xPO4O3} reveals that the incommensurate magnetic structure present in the parent compound ($x=$0) undergoes continuous changes with increasing $x$, until at high enough dilution ($x \sim 0.5$) a spin-glass-like state with no long range order forms.  We find, by analyzing magnetic neutron powder diffraction patterns across the series, that the helical winding length and $ab$ plane correlation length both decrease as $x$ increases, but remain nearly equal to one another over the range $x = 0$ to 0.25.  Based on numerical analysis of a Heisenberg exchange Hamiltonian with competing near neighbor ($J_1$) and next near neighbor ($J_2$) interactions, it is observed that the decreasing pitch length is consistent with a decreasing ratio of $J_2/J_1$, as expected from mean field considerations.  Unexpectedly, the $ab$ plane correlation length decreases in the same manner as the helical pitch length; the result is that despite the increased spacing in $Q$ between satellite peaks for higher $x$, they still cannot be individually resolved.   The coincidence of the two length scales suggests an intrinsic mechanism for the truncation of the the helical spin structure, rather than the nucleation of domain walls at impurities.   These results further validate that magnetic frustration is responsible for the development of the helical magnetic structure in \ce{Fe3PO4O3}, while also providing clues as to the mechanism for the development of small needle-like domains of this structure.  We anticipate that future studies of single crystal samples of \ce{Fe3PO4O3} could elucidate the nature of this short range antiferromagnetic helical structure and its relation to topological spin textures such as helical domain walls, or antiferromagnetic Skyrmions.

\section{Acknowledgments}

The authors acknowledge helpful discussions with M. Gelfand.  This research used resources at the High Flux Isotope Reactor, a DOE Office of Science User Facility operated by the Oak Ridge National Laboratory.  

\appendix

\begin{figure*}
	\centering
	\includegraphics[width=\linewidth]{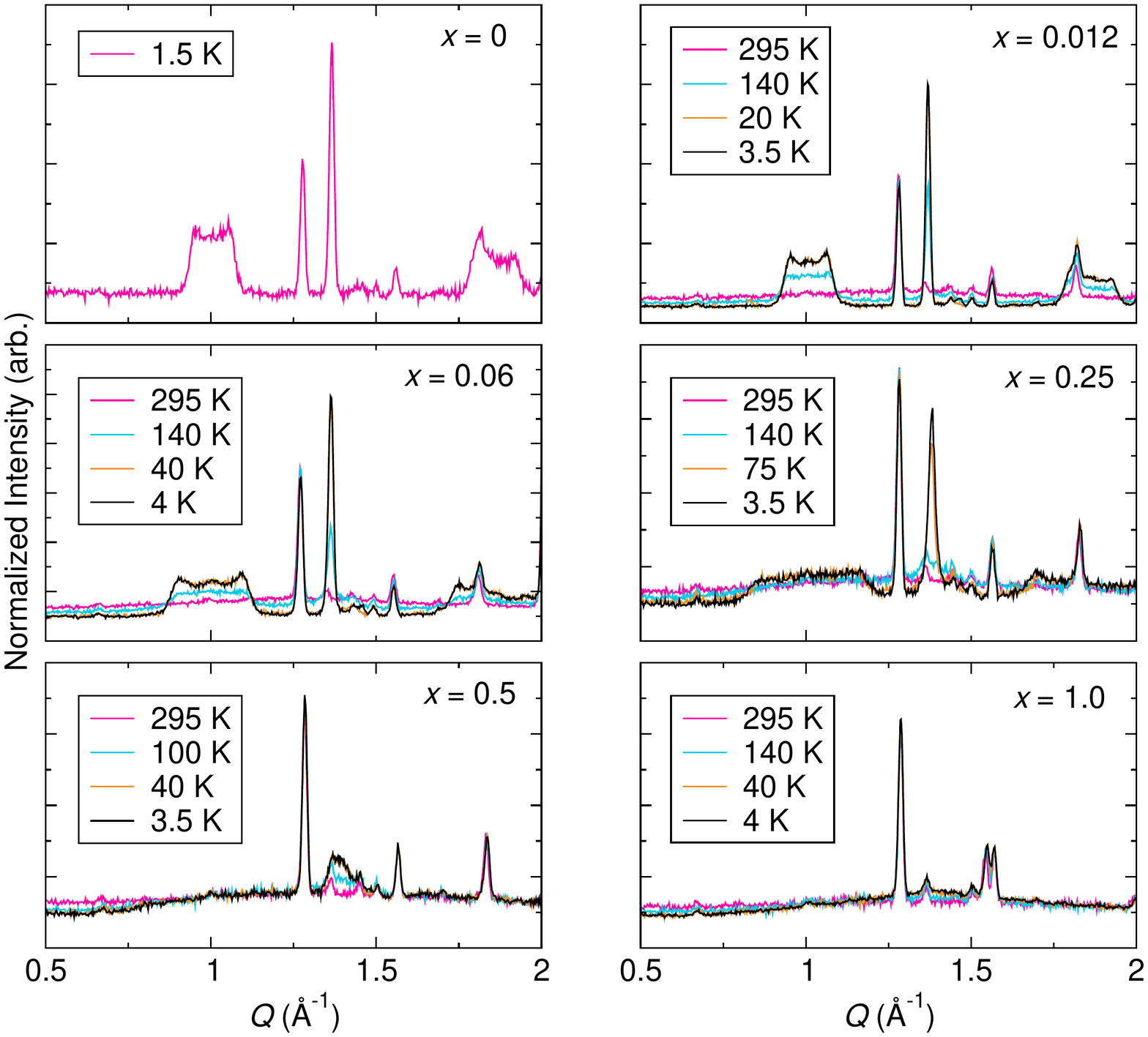}
	\caption{Raw data from neutron powder diffraction of \ce{Fe_{3-x}Ga_xPO4O3} at all temperatures and values of $x$. \ce{FePO4} impurity peaks are denoted with black stars and \ce{GaPO4} impurity peaks are denoted with blue addition signs.}
	\label{fig:neutron_raw}
\end{figure*}

%\bibliography{library}


\end{document}